\documentclass[smallabstract,smallcaptions]{dccpaper}

\usepackage{epsfig}
\usepackage{amsmath}
\usepackage{amssymb}
\usepackage{color}
\usepackage{url}
\usepackage{booktabs}
\usepackage{multirow}
\usepackage{tabularx}
\usepackage{array}

\newlength{\figurewidth}
\newlength{\smallfigurewidth}

\begin{document}

\title{
{\large
\textbf{Semantic-Guided Frequency Modulation\\
for Learned Image Compression}
}
}

\author{%
Gangyi Tian$^{a,c}$, Wen Ji$^{b,a,\ast}$,
Alireza Furutanpey$^{d}$, and Mingxuan Li$^{b}$\\[0.5em]
{\small
\begin{minipage}{\dimexpr\linewidth-2\tabcolsep\relax}
\begin{center}
\begin{tabular}{c}
$^{a}$State Key Laboratory of Processors, Institute of Computing Technology,\\
Chinese Academy of Sciences, Beijing, China\\
$^{b}$Institute of AI for Industries, Chinese Academy of Sciences, Nanjing, China\\
$^{c}$University of Chinese Academy of Sciences, Beijing, China\\
$^{d}$Distributed Systems Group, Vienna University of Technology, Vienna, Austria\\[0.3em]
\url{tiangangyi22b@ict.ac.cn};
\url{wji@iaii.ac.cn};\\
\url{a.furutanpey@dsg.tuwien.ac.at};
\url{mxli@iaii.ac.cn}
\end{tabular}
\end{center}
\end{minipage}}
}

\maketitle
\thispagestyle{empty}

\begin{abstract}
High-frequency attenuation can reduce the rate of learned image codecs but
also remove object boundaries and fine detail. We propose Semantic-Guided
Adaptive Latent-Frequency Modulation (SG-AFM), which uses a semantic
structural prior to guide attenuation of encoder latents. SG-AFM applies
a Haar wavelet transform and combines a fixed semantic-frequency
multiplier with a bounded learned correction to modulate high-frequency
subbands. The semantic prior is computed only at the encoder. No semantic
map is transmitted, and no decoder modules are added. Across six
datasets, SG-AFM reduces PSNR-based BD-rate relative to TCM by an average
of 13.17\%. It achieves the best BD-rate among the compared methods on
ExDark, NightCity, and OnePlus, with a mean reduction of 16.80\%. An
ablation on OnePlus demonstrates the benefit of spatially aligned
semantic guidance, and visual comparisons illustrate improved
reconstruction of local structures.
\end{abstract}

\section{Introduction}
\label{sec:introduction}

In images with complex illumination, object boundaries and fine detail
appear alongside glare, reflections, and sharp exposure transitions. High-frequency latent responses can occur in all
of these regions, so their magnitude alone does not identify which
structures to prioritize during compression. The information a codec
must preserve depends on its specified reconstruction and prediction
requirements~\cite{furutanpey2026taskscope}. Attenuating these responses
can reduce bitrate, but applying attenuation without spatial guidance
can also degrade reconstruction. We use semantic structure to guide
where and how strongly to attenuate the high-frequency components of
encoder latents.

Learned image compression (LIC) methods improve rate-distortion (RD)
performance through learned transforms, entropy models, and quantization.
Balanced RD optimization further improves compression efficiency by
coordinating the competing rate and distortion objectives during
training~\cite{zhang2025balanced}.
TCM~\cite{liu2023tcm} combines convolutional and Transformer architectures
for local and non-local modeling. MambaIC~\cite{zeng2025mambaic} and
Cassic~\cite{qin2025cassic} use state-space models for long-range and
content-adaptive representation learning, while adaptive lattice vector
quantization~\cite{xu2025multirate} supports multirate coding.
Frequency-aware methods explicitly process frequency components.
FTIC~\cite{li2024ftic} combines multiscale directional analysis with
adaptive frequency modulation, and SAAF~\cite{ma2026saaf} combines sparse
attention with content-adaptive frequency processing. Frequency-biased
compression and compensation~\cite{liu2025frequency} couples frequency
suppression with reconstruction.
Semantic information has also been used to refine generative image
compression~\cite{xia2025diffpc}, while region-adaptive latent masking
allocates coding resources according to local image
importance~\cite{xu2025region}.
We introduce an image-derived semantic
prior into latent-frequency modulation to guide the spatial distribution
of attenuation.

\begin{figure}[t]
    \centering
    \includegraphics[width=\linewidth]{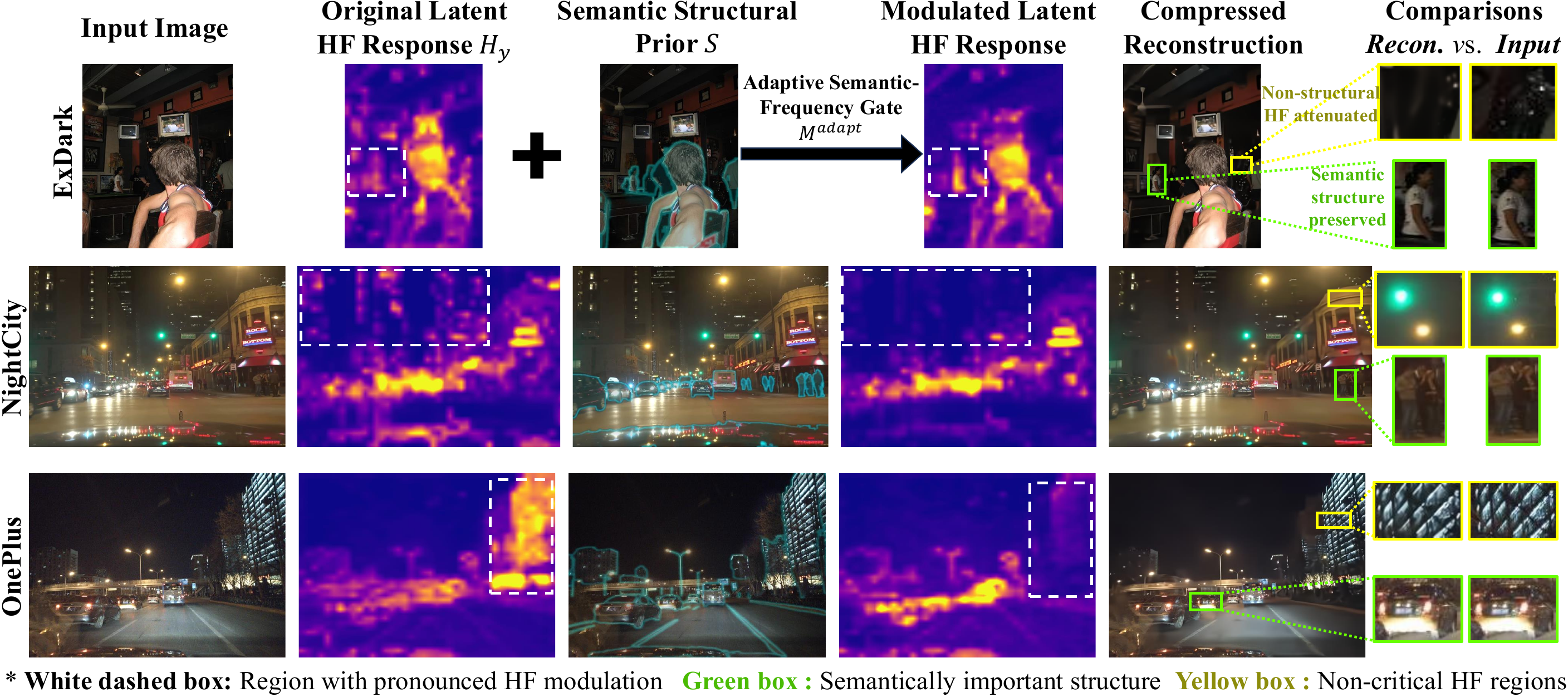}
    \caption{Semantic guidance for latent-frequency modulation. Examples
    from ExDark, NightCity, and OnePlus show the input, original
    high-frequency (HF) response $H_y$, semantic structural prior $S$,
    modulated response, and reconstruction. Green boxes mark structural
    regions and yellow boxes mark other HF regions. White dashed boxes
    locate examples of modulation in the response maps.}
    \label{fig:motivation}
\end{figure}

Figure~\ref{fig:motivation} illustrates the semantic and frequency
information used by our method. The latent high-frequency response
highlights local activations, while a structural prior derived from
SAM~\cite{carion2025sam3} identifies regions to prioritize. We combine
these two inputs in Semantic-Guided Adaptive Latent-Frequency Modulation
(SG-AFM). We use a fixed multiplier to apply less attenuation at locations with
higher semantic importance and learn a residual correction for each
channel and spatial location.

We insert SG-AFM between the analysis and hyper-analysis transforms of
TCM. A discrete wavelet transform (DWT) separates the latent into low-
and high-frequency subbands. We modulate the high-frequency coefficients
and reconstruct the latent before entropy coding. Semantic extraction
and modulation operate only at the encoder, so the decoder requires no
semantic map or additional modules.

Our contributions are:
\begin{itemize}
    \item We introduce semantic-guided modulation of high-frequency
    latent subbands using a fixed semantic-frequency multiplier and a
    bounded learned correction. The module adjusts attenuation across
    channels and spatial locations without transmitting a semantic map
    or changing the decoder architecture.
    \item We evaluate SG-AFM on six datasets and obtain a mean
    PSNR-based BD-rate reduction of 13.17\% relative to TCM. SG-AFM
    achieves the best BD-rate among the compared methods on ExDark,
    NightCity, and OnePlus, averaging 16.80\% lower rate than TCM.
    An ablation on OnePlus demonstrates the benefit of spatially
    aligned semantic guidance over blind and shuffled modulation.
\end{itemize}

\section{Proposed Method}
\label{sec:method}

\subsection{Encoder-Side Semantic Guidance}
\label{sec:overall}

\begin{figure}[h]
    \centering
    \includegraphics[width=\linewidth]{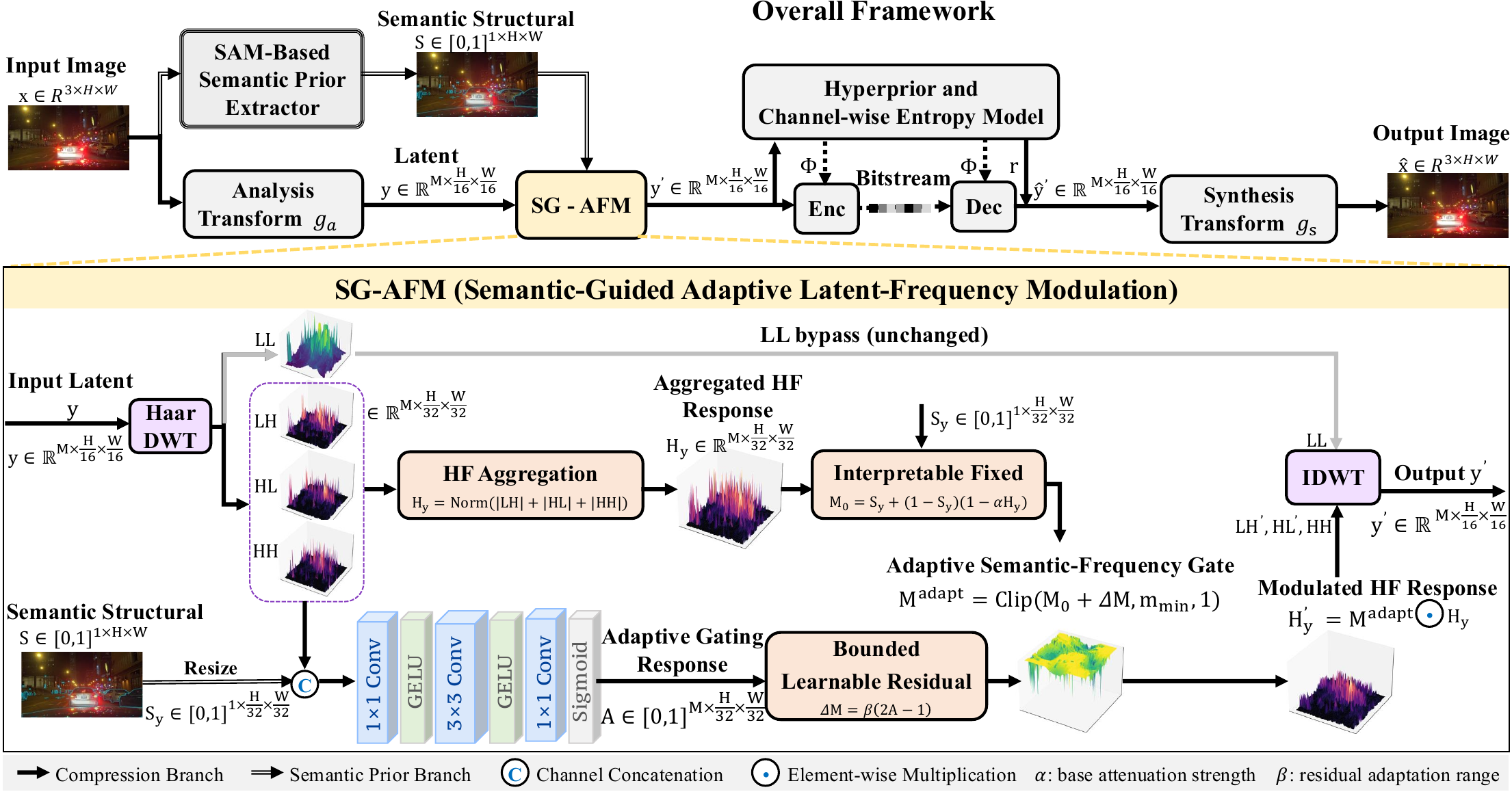}
    \caption{SG-AFM within the TCM codec. We derive the semantic prior
    $S$ from the input image and combine it with wavelet responses of
    the encoder latent $y$. We compute the modulation weights from a fixed multiplier and a bounded
    learned correction. We multiply each
    signed high-frequency subband by these weights and retain the
    low-frequency subband before IDWT reconstruction of $y'$. The
    aggregate $H_y'$ visualizes the modulated response. Entropy coding
    retains the TCM main-latent and hyperlatent streams and transmits no
    semantic map.}
    \label{fig:framework}
\end{figure}

Given an image $x\in\mathbb{R}^{3\times H\times W}$, the TCM analysis
transform $g_a$ produces the latent representation
$y\in\mathbb{R}^{M\times H/16\times W/16}$. We extract object and
structural cues from $x$ using SAM~\cite{carion2025sam3} and represent
them as a semantic structural prior
$S\in[0,1]^{1\times H\times W}$. We assign greater structural importance to locations with larger values
of $S$ when computing the fixed attenuation multiplier. SG-AFM processes $y$ and $S$ to produce a modulated latent
$y'$ with the same dimensions as $y$, as shown in
Fig.~\ref{fig:framework}.

The TCM hyperprior and channel-wise entropy model encode the quantized
modulated latent together with its hyperlatents. The synthesis transform
$g_s$ reconstructs $\hat{x}$ from the entropy-decoded latent $\hat{y}'$.
The decoder does not require $S$. We retain the decoder architecture and
perform all semantic extraction and latent modulation at the encoder.

\subsection{Latent-Frequency Decomposition and Fixed Prior}
\label{sec:frequency}

A one-level Haar DWT decomposes $y$ into a low-frequency subband $LL$
and three directional high-frequency subbands $LH$, $HL$, and $HH$,
each of size $M\times H/32\times W/32$. We retain $LL$ and compute an
aggregate response from the magnitudes of the high-frequency
coefficients,
$H_y=\mathrm{Norm}(|LH|+|HL|+|HH|)$.
Here, $\mathrm{Norm}(\cdot)$ rescales the response to $[0,1]$, giving
$H_y\in[0,1]^{M\times H/32\times W/32}$. Each entry of $H_y$ describes
the high-frequency response at one latent channel and spatial location.

We resize $S$ to the subband resolution and broadcast it along the
channel dimension to obtain
$S_y\in[0,1]^{M\times H/32\times W/32}$. The fixed semantic-frequency
multiplier is
\begin{equation}
M_0=S_y+(1-S_y)(1-\alpha H_y),
\label{eq:fixed_prior}
\end{equation}
where $\alpha$ sets the base attenuation strength. Equivalently,
$M_0=1-\alpha(1-S_y)H_y$. At a fixed high-frequency response, larger
values of $S_y$ give a multiplier closer to one and therefore less
attenuation. At locations with low semantic importance, stronger
high-frequency responses receive more attenuation. The fixed term thus
combines the amount of local high-frequency activity with its assigned
semantic importance.

\subsection{Adaptive Semantic-Frequency Modulation}
\label{sec:adaptive_modulation}

Different directional patterns can produce the same aggregate response
$H_y$. We use the three directional magnitude maps separately to learn
a residual correction that adjusts $M_0$ to the image content and
channel statistics.

We concatenate $|LH|$, $|HL|$, and $|HH|$ with $S_y$ and process them
with a $1\times1$ convolution, a Gaussian Error Linear Unit (GELU), a
$3\times3$ convolution, a second GELU, a $1\times1$ convolution, and a
sigmoid activation. Through the $3\times3$ layer, we can use neighboring
responses as well as directional composition to determine the
attenuation weight shared by the three directional subbands. The
predictor output is
$A\in[0,1]^{M\times H/32\times W/32}$. We compute the final
modulation weights as
\begin{equation}
M^{\mathrm{adapt}}
=
\mathrm{Clip}
\left(
M_0+\beta(2A-1),\,m_{\min},\,1
\right),
\label{eq:adaptive_modulation}
\end{equation}
where $\beta$ sets the range of the residual and $m_{\min}$ is the lower
clipping threshold. The correction $\Delta M=\beta(2A-1)$ lies in
$[-\beta,\beta]$ and adjusts the attenuation separately for each channel
and location. Clipping at one prevents amplification of the
high-frequency coefficients.

We use $\alpha=0.02$, $\beta=0.05$, and $m_{\min}=0.90$ for every
dataset and rate point. These values imply $M_0\in[0.98,1]$ and
$M^{\mathrm{adapt}}\in[0.93,1]$, so the lower clipping threshold is
inactive in this configuration. The learned correction can increase or
decrease attenuation relative to the semantic preference in $M_0$.

\subsection{High-Frequency Modulation and Coding}
\label{sec:coding}

We apply the same weights to the three signed high-frequency subbands
and retain $LL$,
\begin{equation}
\begin{aligned}
LH'&=M^{\mathrm{adapt}}\odot LH,\quad
HL'=M^{\mathrm{adapt}}\odot HL,\\
HH'&=M^{\mathrm{adapt}}\odot HH,\quad
y'=\mathrm{IDWT}(LL,LH',HL',HH'),
\end{aligned}
\label{eq:modulation}
\end{equation}
where $\odot$ denotes element-wise multiplication and IDWT is the inverse
wavelet transform. The resulting $y'$ has the original latent dimensions
and enters the TCM hyper-analysis and entropy-coding path. The bitstream
contains the coded main latent and hyperlatents. No additional stream is
needed for $S$ or the modulation weights. For visualization, we define
$H_y'=M^{\mathrm{adapt}}\odot H_y$ separately from the signed subbands
used in the inverse transform.

\section{Experimental Results}
\label{sec:experiments}

\subsection{Datasets and Experimental Settings}
\label{sec:settings}

We evaluate SG-AFM on six datasets. Kodak~\cite{kodak_dataset} and
Tecnick~\cite{asuni2015testimages} are normal-light compression
benchmarks. ExDark~\cite{loh2019exdark},
NightCity~\cite{tan2021nightcity}, and
OnePlus~\cite{Yu2021ReconfigISPRC} contain complex-illumination images,
including nighttime scenes and uneven exposure. We also evaluate on the
low-light VELoL dataset~\cite{liu2021velol}. We include every image in
each evaluation set without content-based selection.

We use TCM~\cite{liu2023tcm} as the compression reference and compare
with BPG~\cite{bellard2018bpg}, VTM-23.3 Intra~\cite{jvet_vtm233},
Cheng-attn~\cite{cheng2020learned}, TCM-AuxT~\cite{li2025on},
DCAE~\cite{lu2025learned}, and LALIC~\cite{Feng2025LALIC}. All
experiments use a single NVIDIA Tesla V100 GPU. The modulation
parameters are identical across datasets and rate points. For all
modulation variants in the ablation, we jointly fine-tune the encoder
and synthesis transform while retaining the synthesis architecture.

We measure rate in bits per pixel (bpp) and reconstruction quality using
peak signal-to-noise ratio (PSNR) and multi-scale structural similarity
(MS-SSIM). PSNR is reported in decibels (dB), and MS-SSIM is converted
to the logarithmic scale as
$-10\log_{10}(1-\mathrm{MS\text{-}SSIM})$. PSNR-based Bjøntegaard delta rate (BD-rate)
compares rate with the original TCM at equivalent quality. Negative
values indicate rate savings. We compute group averages as arithmetic
means of the dataset BD-rate values. For semantic regions of interest
(RoIs), RoI-PSNR measures reconstruction fidelity, RoI-L1 measures
pixel-wise error, and RoI-Grad measures gradient-domain error. Higher
PSNR and MS-SSIM are better, while lower RoI-L1 and RoI-Grad are better.

\subsection{Rate-Distortion Performance}
\label{sec:rd_results}

\begin{figure}[t]
    \centering
    \includegraphics[width=\linewidth]{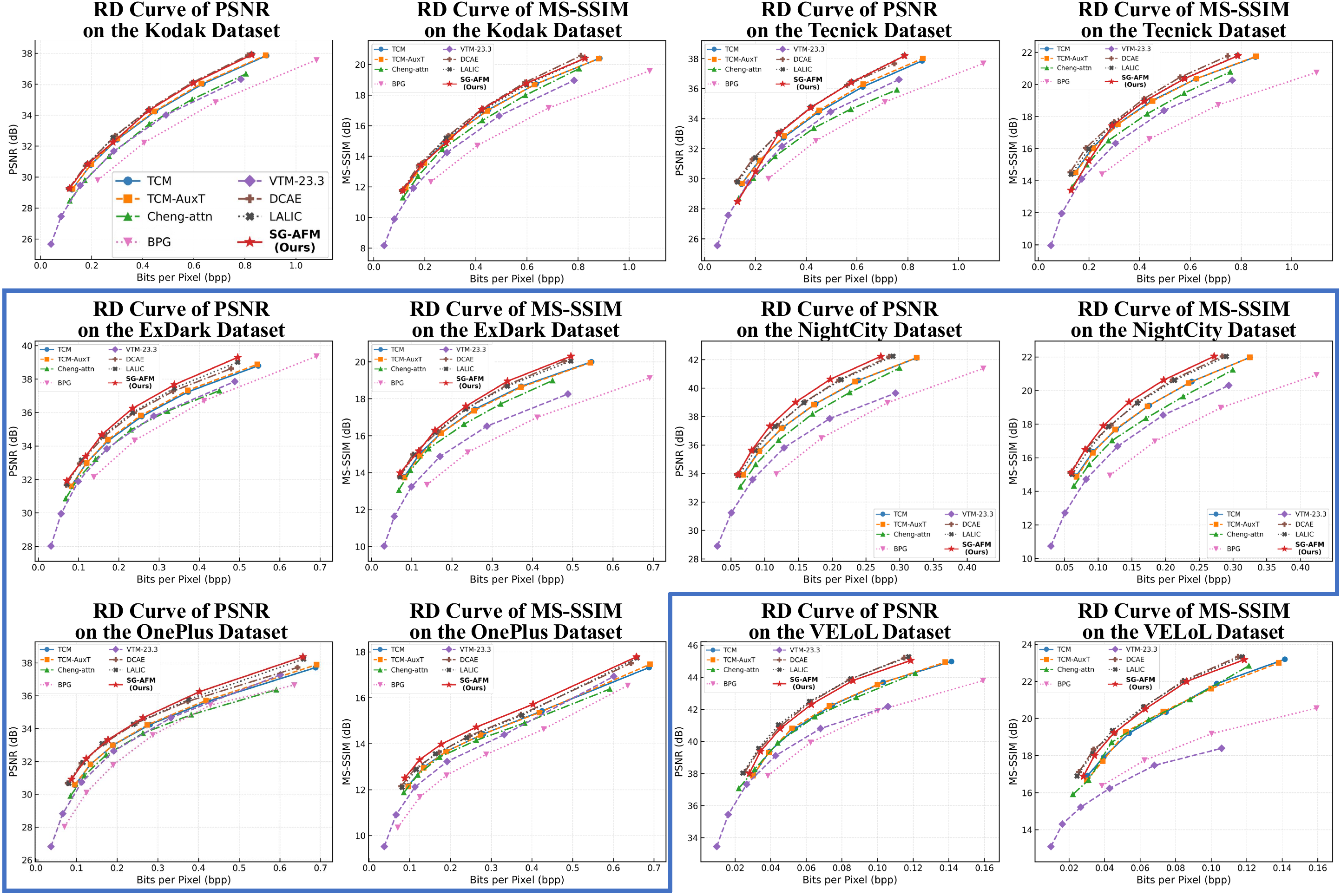}
    \caption{Rate-distortion curves on six datasets, measured by PSNR
    and MS-SSIM against bits per pixel. Blue boxes highlight ExDark,
    NightCity, and OnePlus, the three complex-illumination benchmarks.
    Kodak and Tecnick provide normal-light comparisons, and VELoL
    provides a low-light comparison.}
    \label{fig:rd}
\end{figure}

SG-AFM reduces PSNR-based BD-rate relative to TCM on all six datasets,
with an average reduction of $13.17\%$. Table~\ref{tab:bdrate} gives
the dataset comparisons, and Fig.~\ref{fig:rd} shows the PSNR and
MS-SSIM curves. SG-AFM achieves the best BD-rate among the compared
methods on all three complex-illumination datasets. The reductions
relative to TCM are $17.06\%$ on ExDark, $16.25\%$ on NightCity, and
$17.10\%$ on OnePlus.

\begin{table}[!t]
\centering
\caption{PSNR-based BD-rate (\%) relative to TCM. Negative values
indicate rate savings. More negative values are better. N-Avg. and
C-Avg. denote averages over the normal-light and complex-illumination
datasets, respectively. Best results are highlighted in bold.}
\label{tab:bdrate}

\scriptsize
\renewcommand{\arraystretch}{1.05}
\setlength{\tabcolsep}{2pt}

\begin{tabularx}{\textwidth}{
@{}
l
*{8}{>{\centering\arraybackslash}X}
@{}
}
\toprule

\multirow{2}{*}{Method}
& \multicolumn{2}{c}{Normal Light}
& \multicolumn{3}{c}{Complex Illumination}
& \multicolumn{1}{c}{Low Light}
& \multicolumn{2}{c}{Average} \\

\cmidrule(lr){2-3}
\cmidrule(lr){4-6}
\cmidrule(lr){7-7}
\cmidrule(lr){8-9}

&
Kodak &
Tecnick &
ExDark &
NightCity &
OnePlus &
VELoL &
N-Avg. &
C-Avg. \\

\midrule

BPG~\cite{bellard2018bpg}
& +37.84
& +42.59
& +31.51
& +61.81
& +21.01
& +42.60
& +40.22
& +38.11 \\

VTM-23.3~\cite{jvet_vtm233}
& +16.09
& +11.63
& +13.37
& +36.09
& +6.23
& +21.00
& +13.86
& +18.56 \\

\midrule

Cheng-attn~\cite{cheng2020learned}
& +14.73
& +20.53
& +11.86
& +11.97
& +10.66
& +1.63
& +17.63
& +11.50 \\

TCM-AuxT~\cite{li2025on}
& -0.87
& -1.94
& -0.38
& -0.04
& -1.52
& -0.97
& -1.41
& -0.65 \\

DCAE~\cite{lu2025learned}
& \textbf{-8.63}
& \textbf{-13.24}
& -12.30
& -11.60
& -13.97
& -19.26
& \textbf{-10.94}
& -12.62 \\

LALIC~\cite{Feng2025LALIC}
& -7.62
& -12.75
& -14.50
& -9.99
& -15.32
& \textbf{-20.86}
& -10.19
& -13.27 \\

\textbf{SG-AFM (Ours)}
& -6.14
& -6.88
& \textbf{-17.06}
& \textbf{-16.25}
& \textbf{-17.10}
& -15.58
& -6.51
& \textbf{-16.80} \\

\bottomrule
\end{tabularx}
\end{table}

Across the three complex-illumination datasets, SG-AFM averages a
$16.80\%$ BD-rate reduction, exceeding LALIC's mean reduction of
$13.27\%$ by $3.53$ percentage points. LALIC is the strongest competing
method by this group average. On NightCity, SG-AFM achieves
$-16.25\%$, compared with $-11.60\%$ for DCAE and $-9.99\%$ for LALIC.
SG-AFM improves the rate-distortion tradeoff over TCM in both PSNR and
MS-SSIM on these three datasets (Fig.~\ref{fig:rd}).

SG-AFM also improves on TCM outside this group. It reduces BD-rate by
$6.14\%$ on Kodak and $6.88\%$ on Tecnick, averaging $6.51\%$ on the
normal-light benchmarks. On VELoL, the reduction is $15.58\%$. DCAE and
LALIC achieve larger reductions on these three datasets. SG-AFM's
strongest comparative performance is on ExDark, NightCity, and OnePlus,
where it ranks first among the evaluated methods.

\subsection{Effect of Semantic Guidance}
\label{sec:ablation}

We compare aligned semantic guidance with two alternative modulation
variants on OnePlus at $\lambda=0.0483$. The blind variant applies
frequency modulation without semantic guidance. Random-shuffle uses a
spatially shuffled prior to disrupt its alignment with the latent.
Table~\ref{tab:ablation} reports the full-image and region-wise metrics.

\begin{table}[h]
\centering
\caption{Ablation study on OnePlus at $\lambda=0.0483$. All modulation
variants jointly fine-tune the synthesis transform without changing its
architecture. RoI denotes the semantic region of interest. Best results
are highlighted in bold.}
\label{tab:ablation}

\footnotesize
\renewcommand{\arraystretch}{1.05}
\setlength{\tabcolsep}{2.0pt}

\begin{tabularx}{\textwidth}{
@{}
>{\raggedright\arraybackslash}p{0.16\textwidth}
>{\centering\arraybackslash}X
>{\centering\arraybackslash}p{0.16\textwidth}
>{\centering\arraybackslash}X
>{\centering\arraybackslash}X
>{\centering\arraybackslash}X
>{\centering\arraybackslash}X
@{}
}
\toprule

Variant &
PSNR$\uparrow$ &
MS-SSIM(dB)$\uparrow$ &
bpp$\downarrow$ &
RoI-PSNR$\uparrow$ &
RoI-L1$\downarrow$ &
RoI-Grad$\downarrow$ \\

\midrule

TCM Anchor
& 37.7183
& 17.3199
& 0.687595
& 36.6200
& 0.011387
& 0.046202 \\

Blind
& 37.6466
& 17.2660
& \textbf{0.654422}
& 36.5456
& 0.011484
& 0.046752 \\

Random-shuffle
& 37.6400
& 17.2701
& 0.657338
& 36.5604
& 0.011477
& 0.046650 \\

\textbf{ Ours}
& \textbf{38.4718}
& \textbf{17.8835}
& 0.656493
& \textbf{36.6994}
& \textbf{0.011322}
& \textbf{0.046068} \\

\bottomrule
\end{tabularx}
\end{table}

With aligned semantic guidance, SG-AFM achieves $38.4718$\,dB PSNR at
$0.656493$ bpp. It improves PSNR by $0.7535$\,dB and reduces rate by
$4.52\%$ relative to the TCM reference. Blind modulation reaches the
lowest rate, $0.654422$ bpp, but yields $37.6466$\,dB PSNR. Using
aligned guidance therefore improves PSNR by $0.8252$\,dB over blind
modulation at $0.32\%$ higher bpp.

The shuffled prior gives $37.6400$\,dB PSNR at $0.657338$ bpp. Aligned
guidance improves both rate and reconstruction quality relative to this
variant. SG-AFM also achieves the best MS-SSIM and all three RoI
metrics in the table. Blind and shuffled modulation reduce bitrate but
lower PSNR relative to TCM. With aligned semantic guidance, SG-AFM
achieves both lower bitrate and higher PSNR than TCM. This comparison
demonstrates the value of spatial
alignment when using semantic information to guide latent-frequency
attenuation.

\subsection{Visual and Region-Wise Quality}
\label{sec:subjective}

\begin{figure}[ht]
    \centering
    \includegraphics[width=0.84\linewidth]{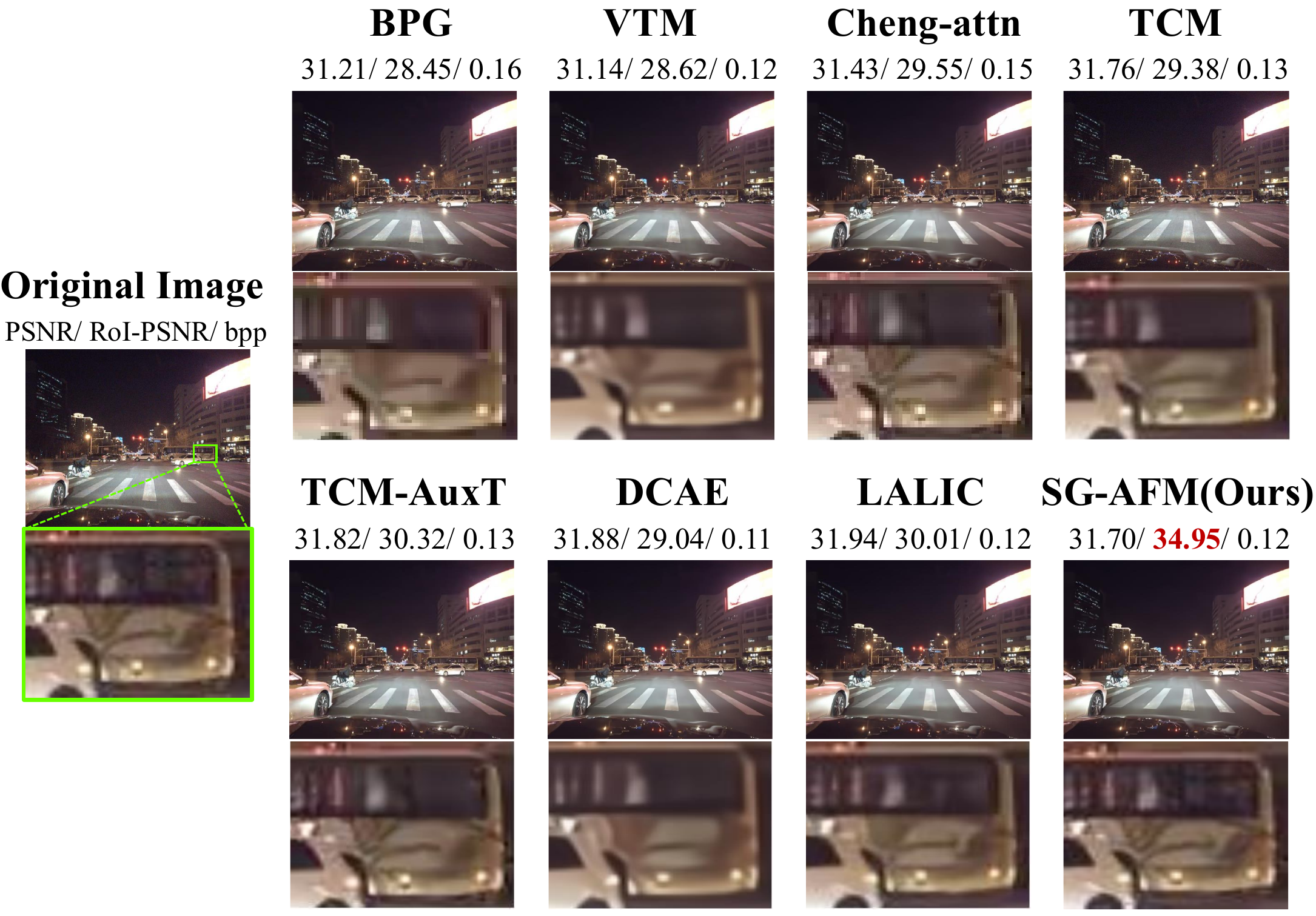}
    \caption{Visual and region-wise reconstruction quality in a
    complex-illumination scene. For each method, the reconstructed image
    is followed by an enlargement of the highlighted structural region.
    Labels report PSNR/RoI-PSNR/bpp. SG-AFM preserves more structural
    detail in the enlarged region at a comparable rate.}
    \label{fig:qualitative}
\end{figure}

Figure~\ref{fig:qualitative} compares reconstructions of a
complex-illumination scene and enlarges a structural region. SG-AFM
preserves the local contours and boundary details, whereas the
competing methods exhibit stronger smoothing in the enlarged region.
At approximately $0.12$ bpp, it achieves $34.95$\,dB RoI-PSNR, compared
with $30.01$\,dB for LALIC and $30.32$\,dB for TCM-AuxT.

Relative to LALIC at the displayed $0.12$ bpp, SG-AFM increases
RoI-PSNR by $4.94$\,dB. Its full-image PSNR is $31.70$\,dB, compared
with $31.94$\,dB for LALIC. The example illustrates improved fidelity
within the semantic region alongside a $0.24$\,dB reduction in
full-image PSNR.

\section{Conclusion}
\label{sec:conclusion}

SG-AFM uses a semantic structural prior to guide attenuation of
high-frequency latent subbands before entropy coding. We combine a
fixed semantic-frequency multiplier with a bounded learned correction
and retain the TCM decoder architecture without transmitting a semantic
map. SG-AFM reduces PSNR-based BD-rate relative to TCM on six datasets,
with a mean reduction of $13.17\%$. It achieves the best BD-rate among
the compared methods on ExDark, NightCity, and OnePlus, averaging
$16.80\%$ lower rate than TCM. The OnePlus ablation demonstrates that
spatially aligned semantic guidance improves the RD performance of
latent-frequency modulation, and the visual example illustrates its
benefit for local structural reconstruction.

\section{Acknowledgements}
\label{sec:acknowledgements}
This work was supported by the National Key Research and Development
Program of China (No. 2024YFE0212000) and the Jiangsu Key Laboratory of
Industrial AI. The corresponding author is Wen Ji.

\Section{References}
\normalsize
\bibliographystyle{IEEEtran}
\bibliography{refs}

\end{document}